\documentclass[usenatbib,usegraphicx,twocolumn]{mn2e}

\def\mnras{MNRAS}
\def\apj{Ap.J}

\def\u{{\bf U}}
\def\th{\vec{\theta}}

\def\del{\partial}
\usepackage{epsfig}
\usepackage{graphics}
\begin{document}
\title[Foregrounds for $21 \, {\rm cm}$: GMRT  observations.]
{Foregrounds for redshifted $21 \, {\rm cm}$ studies of reionization:  
 GMRT $153 \,{\rm MHz}$  observations.}
\author[ S. S. Ali, S. Bharadwaj and J. N. Chengalur ]{ 
  Sk. Saiyad Ali$^{1}$\thanks{Email:saiyad@cts.iitkgp.ernet.in} 
  Somnath
  Bharadwaj$^{1}$\thanks{Email:somnath@cts.iitkgp.ernet.in},
  and Jayaram N. Chengalur $^{2}$\thanks{Email:chengalu@ncra.tifr.res.in}
  \\ $^{1}$ Department of Physics and Meteorology \&
  Centre for Theoretical Studies , IIT Kharagpur,  721 302 , 
  India 
  \\$^2$ National Centre for Radio Astrophysics, TIFR, Post Bag 3,
  Ganeshkhind, Pune 411 007, India} 
\maketitle

\begin{abstract} 

Foreground subtraction is the biggest challenge for  future 
redshifted $21 \, {\rm cm}$ observations  to probe reionization.    
We use a short  GMRT observation at $153 \, {\rm  MHz}$
to  characterize the statistical properties of the background
radiation across  $\sim1^{\circ}$ to sub-arcminutes  angular scales,
and across  a  frequency band of $5 \, {\rm MHz}$ with $62.5 \, {\rm 
kHz}$ resolution. The statistic we use is the visibility correlation
function, or equivalently the angular power spectrum $C_l$.
We present the results obtained from using relatively unsophisticated, 
conventional data calibration procedures. We find that even
fairly simple minded calibration allows one to estimate
the visibility correlation function at a given frequency
$V_2(U,0)$.  From our observations we find that $V_2(U,0)$ is
consistent with foreground model predictions at all angular scales except
the largest ones probed by our observations where the 
the model predictions are somewhat in excess. 
 On the other hand
the  visibility correlation between  different frequencies 
$\kappa(U,\Delta \nu)$, seems to be much more sensitive to calibration
errors. We find a rapid decline in $\kappa(U,\Delta \nu)$, in contrast 
with the prediction  of less than $1 \%$ variation across  $2.5 \, {\rm MHz}$.
In this case however, it seems likely that a substantial part
of the discrepancy may be due to limitations of data reduction
procedures.

\end{abstract}


\section{Introduction}
Observations of  redshifted $ 21 {\rm cm}$ radiation from the
 large scale distribution of neutral  hydrogen (HI) are perceived as
 one of the most promising future probes of the Universe at high
 redshifts (see \citealt{furla} for a recent review).
Observational evidence from  quasar absorption spectra
 \citep{becker,fan} and the CMBR  \citep{spergel,page}
 together imply that the HI was reionized  over an extended period
 spanning the redshift range $6 \le z \le 15$ (for reviews see
 \citealt{barkana,fck06,cf06}).  Determining how and when the Universe was
 reionized is one of the most important issues that will be addressed
 by future $21 {\rm cm}$ observations. The Giant Meter Wave  Radio
 Telescope  (GMRT  \footnote{http://www.gmrt.ncra.tifr.res.in};  
 \citealt{swarup}), currently  functioning at several frequency bands in
the  range $150$ to $ 1420 \, {\rm MHz}$ is very well suited for
 carrying out initial investigations towards detecting the
 reionization HI signal.  There are several upcoming    low-frequency
instruments  such  as   LOFAR\footnote{http://www.lofar.org/},
 MWA\footnote{http://www.haystack.mit.edu/arrays/MWA},
 21CMA\footnote{http://web.phys.cmu.edu/$\sim$past/} 
 and   SKA\footnote{http://www.skatelescope.org/} which are being
 built specifically with these observations in view. 

It is currently perceived that a statistical analysis of the
fluctuations in the redshifted $21 \, {\rm cm}$ signal holds the
greatest potential for observing   HI at high redshifts
\citep{BS1,zald,morales,BA5,BP5}. Correlations among the visibilities
measured in radio-interferometric observations directly probe the
HI power spectrum  at the epoch where the radiation originated. 
The reionization visibility signal at the GMRT is expected to be $\sim
1 \,{\rm mJy}$ 
and smaller (\citealt {BA5}).  This HI signal is present as a minute
component of the background in all low frequency observations, and it
is buried in foreground radiation from other astrophysical sources
whose contribution is $4$ to $5$ orders of magnitude
larger. Extracting the HI signal from the foregrounds is a major
challenge. 

 Individual sources can be  identified and removed from
the image at a flux level which depends on the sensitivity. 
The contribution from the remaining discrete sources could be
large enough to overwhelm the  HI signal \citep{dmat1}.  The  
diffuse synchrotron emission from
our Galaxy \citep{shaver} is  another  important component. 
 Foreground sources include  free-free  emission from ionizing
halos  \citep{oh}, faint   radio loud quasars \citep{dmat1} and
synchrotron emission from low redshift galaxy clusters
\citep{dmat}.  

The foregrounds are expected to have a continuum spectra, and the
contribution at two different frequencies separated by $\Delta \nu
\sim 1 \,{\rm MHz}$ are expected to be highly correlated. The HI
signal is expected to be  uncorrelated at such a frequency 
separation and this holds the promise of allowing us to separate the
signal from the foregrounds. A possible approach is to subtract a best
fit continuum spectra for each line of sight \citep{wang}
and then
use the residuals to determine the HI power spectrum. An alternate
approach is to first determine the statistical properties of the total
radiation and then subtract out the   smooth  $\Delta \nu$ dependent
part to extract the HI signal \citep{zald}. The issue of foreground
removal has also been studied by \citet{mor2} and
    \citet{mcquinn}.

It is crucial to accurately characterize the foregrounds in order to
be able to detect the HI signal  in future observations. In this paper 
we used  GMRT observations to characterize the foregrounds  at 
 $153 \, {\rm MHz}$ which corresponds to an HI signal from $z=8.3$. 
 To the best of our knowledge 
this is the first attempt to directly characterize the foregrounds at
angular scales ($\sim 1^{\circ}$ to sub-arcminute) and frequency
coverage  ($6 \, {\rm MHz}$ with $62.5 \, {\rm kHz}$ resolution)
relevant for detecting the  reionization HI signal.  

We next present a brief outline of the paper. 
In Section 2 we describe the observations and data reduction while in 
Section 3   we present ``visibility-correlations'' which we use to
quantify the statistical properties of our radio-interferometric
data.  Section 4 presents the predictions of existing foreground
models, and in Section 5 we present our results and discuss their 
implications. 

\section{GMRT Observations and Data Reduction}
\begin{figure}
\includegraphics[width=75mm]{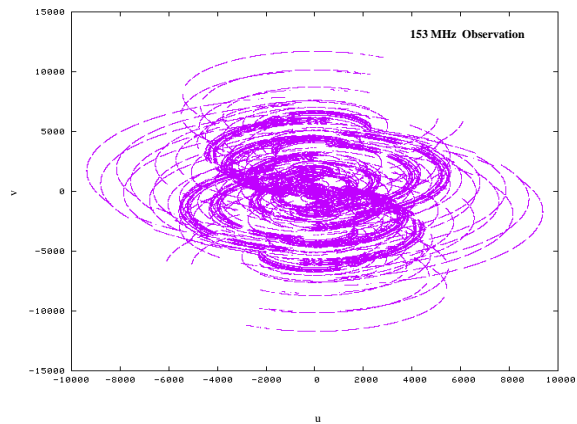}
\caption{This shows the $uv$ coverage of the GMRT data that we have
  analyzed. Here $(u,v)$ are the antenna separations in   wavelength 
  units at the  observing frequency $153 \, {\rm MHz}$.} 
\label{fig:a1}
\end{figure}

The GMRT has a hybrid configuration (\citealt{swarup}) where 
$14$ of the $30$ antennas  are randomly distributed in a Central
 Square $\sim \rm 1.1 \,  km \times 1.1\,  km$ in extent. These provide the
$uv$ coverage at small baselines. Here baseline  refers to the 
 antenna separation, and we use the two dimensional vector $\u$ to
 denote the component perpendicular to the direction of
 observation. Note that $\u$ has Cartesian components $(u,v)$ and 
 is dimensionless being  in units of the observing wavelength.   
The shortest baseline at the GMRT is $\rm 100 \,
 \,   m$  which comes down  to around  $\rm 60 \, m$  with projection effects. 
The rest of the antennas in the GMRT  lie along three arms in an
approximately  'Y' configuration. These provide $uv$ coverage at long
baselines (the  longest  baseline is $\rm 26 \,km$).  
The diameter of each GMRT antenna is $\rm 45 m$. The hybrid
configuration gives reasonably   good sensitivity for both compact and
extended  sources. Figure \ref{fig:a1} shows the $uv$ coverage of our
GMRT observations. 

On $15^{\rm th}$ June, $2005$ we observed a field centered on  
Upsilon Andromedae  (which is an extra-solar  planetary system 
system    at $\alpha_{2000}=1^h36^m48^s,\delta_{2000}=
41^{\circ}24^{'}23^{''})\, $ for a total of 14 hours (including
calibration). No emission that could be attributed to the 
planet was detected in  our observations. The galactic co-ordinates
are $l=132.00^{o}, b = 20.67^{o}$. From the 408 MHz \cite{haslam} map
the sky temperature at this location is $\sim 30$~K (at 408~MHz), and
there is no structure visible at the angular resolution of the map.

The observational
set up used a total of $128$ frequency channels spanning $8 \, {\rm MHz}$ 
centered at  $153 \, {\rm MHz}$. Each frequency channel is $62.5 {\rm kHz}$ 
wide. A  $6 \, {\rm MHz}$ wide band-pass filter was introduced in the IF 
stage to exclude known strong  Radio Frequency Interference (RFI), hence
only $3/4$ of the central channels contain astronomical signals. The  
integration  time was  16 seconds, and  visibilities were recorded for 
two orthogonal circular polarizations. The visibility data were analyzed 
using the  Astronomical Image Processing Software (AIPS). The calibrator
source 3C48 was  used for flux, phase  and bandpass calibration. The 
calibrator was observed every half hour so as to correct for temporal 
variations in the system gain. Standard AIPS tasks were used to flag
all data that could be visually identified as being bad.  We then made a high
resolution  image of the source using only a  single  channel (channel 35). 
The synthesized beam has a FWHM of $29'' \times 25''$ and the rms. noise 
in the CLEANed image is $9.5 \, {\rm mJy/Beam}$.  All sources with flux 
density more than $30 \,{\rm mJy}$ were fitted with clean components (CC),
these components were merged and the visibilities corresponding to 
components  with flux more than $8.6 \, {\rm mJy}$   were subtracted 
from the  multi-channel $uv$ data using UVSUB. The value 
$8.6 \, {\rm mJy}$ was chosen because we find predominantly positive
clean components above this flux level whereas  positive and negative
components are  equally abundant below this. The resultant $uv$  data 
is now expected to be dominated by  noise and residual RFI,  since 
the majority of the point sources have  been removed. Visually inspecting 
the data using the  AIPS  tasks VPLOT and UVHGM, we decided to clip 
the data at $12 \, {\rm Jy}$ whereby visibilities with amplitude greater
than $12 \, {\rm Jy}$ were discarded. The clipping  amplitude is in
principle crucial since one would like to ensure that all baselines 
with  RFI contributions  have been discarded, without throwing away 
any good baselines. In practice we found that the exact clipping
value does not make a substantial difference in our subsequent
analysis. After  this we added back the visibilities corresponding
to all the  CC components that we had subtracted.  To first order,
one could expected that at this stage all strong RFI has been 
removed.

\begin{figure*}
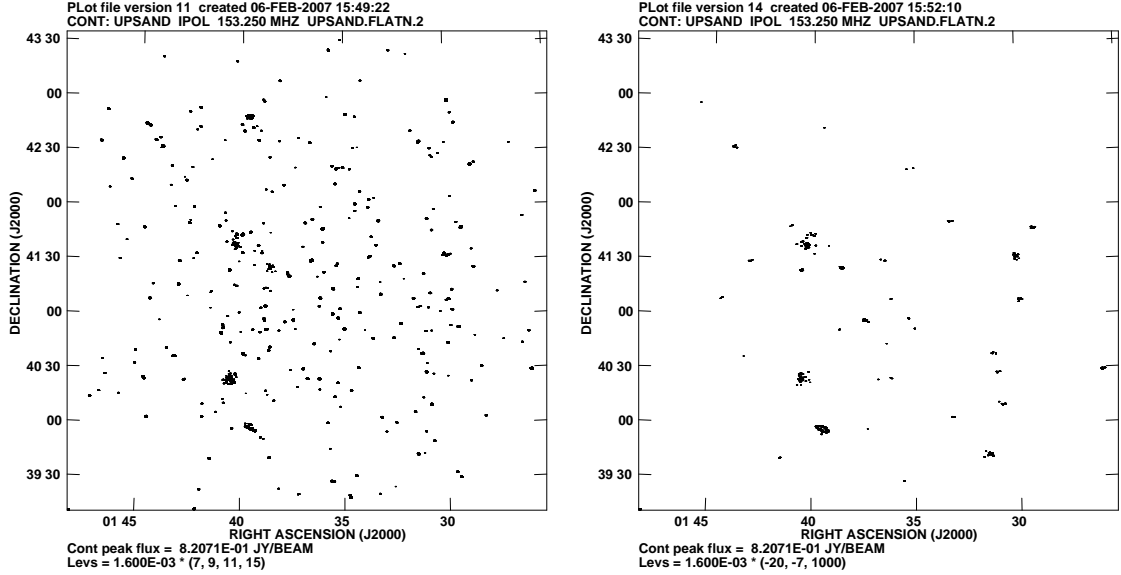

\includegraphics[width=75mm]{beforeuvsub7sigma.ps}
\includegraphics[width=75mm]{beforeuvsubm7sigma.ps}
\caption{These shows our continuum image of bandwidth $5 \, {\rm MHz}$
  centered at $153 \, {\rm MHz}$. The $  4^{\circ}\times 4^{\circ} $
  field was imaged using $139$ facets which have been combined 
  using the AIPS task FLATN.  The rms noise is  $ 1.6 {\rm
  mJy/Beam}$. The left and right panels shows positive and negative
 $ 7-\sigma$ contours respectively. Note that many of the extended
  positive features and all negative features are imaging artifacts
  around the brightest sources.}  
\label{fig:a2}
\end{figure*}

 The large field of view  ($\theta_{\rm FWHM}=3.8^{\circ}$) 
of the GMRT  at $150\, {\rm  MHz}$ lead to considerable errors
if the  non-planar nature  of the GMRT antenna distribution 
is not taken into account.  
We use the three dimensional (3D) imaging feature (e.g.,\citealt{perley})
in the AIPS task IMAGR in which the entire field of view is 
divided into multiple subfields (facets) each of which is imaged
separately. Here a $ 4^{\circ}\times 4^{\circ}$ field of view was imaged
using $139$ facets. We first collapsed $10$ adjacent channels
(channels $30$ to $39$) to make a  single  channel which was used 
to make a CLEANed  image.  This channel's frequency
width $0.625\,  {\rm MHz}$ ($\le 0.7 \, {\rm MHz}$) which is
sufficiently small so as to avoid  bandwidth smearing.
The synthesized beam has FWHM $\sim 20''$
and the cleaned image has rms noise $4.6 \, {\rm mJy/Beam}$. 
The presence of a large number of sources in the field allows
us to do   self calibration loops  to improve the image
quality. The data went through  $4$ rounds of phase self calibration
and a $5$ th round where self calibration was done for both amplitude
and phase. The time interval for the gain correction was chosen as 
$5, 5, 2, 2$ and $2$ \,minutes  for the successive self calibration
loops. The rms. noise in the final cleaned image was $3.1 \, {\rm
  mJy/Beam}$ and the image quality had improved considerably. 
The final gain table was applied to all $128$ frequency channels. 
Channels $21$ to $100$ of this data were then collapsed  into 
8 channels, each containing $10$ of the original frequency 
channels. We use these to  make a  continuum image of the entire 
field. Some more data was flagged at this stage, and we then applied
a final phase self calibration loop. This calibrated data was used
to make the final cleaned  image which is shown in Figure \ref{fig:a2}.
The synthesized beam has a FWHM of  $28'' \times 23''$, and an 
off-source RMS. noise level of $1.6 \, {\rm mJy/Beam}$.  Note 
that several of the extended features like the one at 
$\alpha_{2000}=01^{h} 41^{m}, \delta_{2000}=40^{\circ} 24^{'}$
are actually imaging artifacts around the brightest point sources. 
The brightest  sources  are also found to be accompanied by a 
region of negative flux density, these are presumably the results
if residual phase errors which were not corrected for in our
self calibration process. The maximum and minimum flux density 
in the final image are $820 \,{\rm   mJy/Beam}$ and $-44 \,
{\rm  mJy/Beam}$ respectively. 

\begin{figure*}
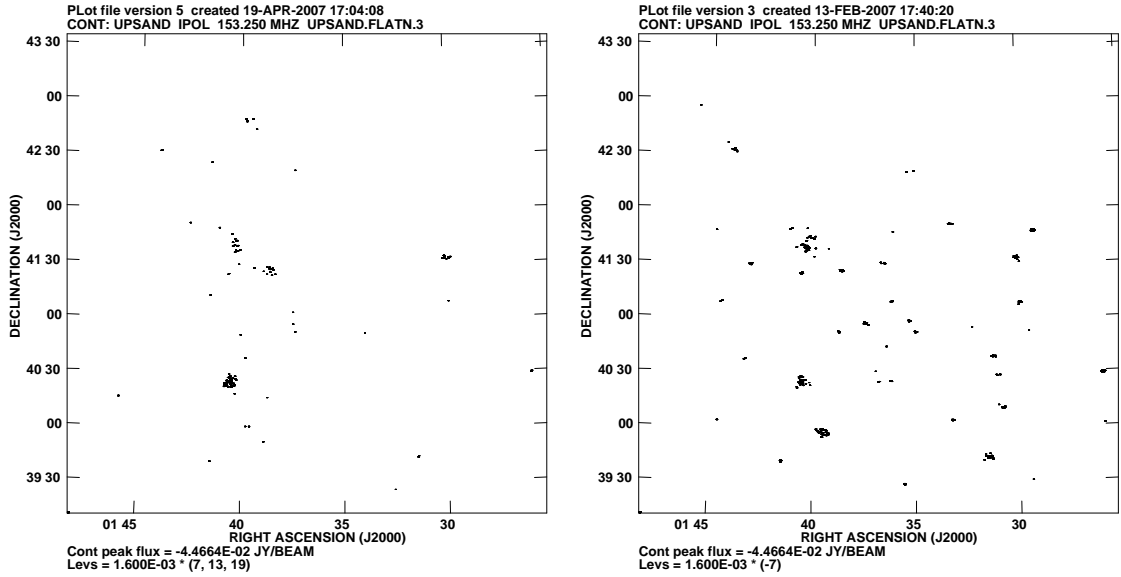

 \includegraphics[width=75.0mm]{afteruvsubp7sigma.ps}
 \includegraphics[width=75.0mm]{afteruvsub7sigma.ps}
  \caption{This is the same as the  Figure \ref{fig:a2} except that
  all the    bright pixels $> 8 \, {\rm mJy/Beam}$  that  were visually
  identified as being genuine sources and not artifacts have been
  fitted   with clean components and removed from the visibility data
 from which  this image  was made. It is expected that most of the
  genuine sources have been removed from this data.}
\label{fig:a3}
\end{figure*}

 Recall that for this experiment, the sources visible in 
the final continuum image (Figure \ref{fig:a2}) are contaminants 
which have to be removed. Pixels with flux density above $8 \, {\rm
mJy/Beam}$ which were visually identified as sources and not imaging
artifacts were fitted with clean components. The clean components were
merged and the visibilities corresponding to these clean components
were subtracted from the original full frequency resolution $uv$ 
data using the AIPS task UVSUB. It is expected that at this stage
most of the genuine sources  in Figure \ref{fig:a2} have been 
removed from the data.  Figure \ref{fig:a3}  shows the final 
image made from the residual visibility data after UVSUB. The 
maximum and minimum flux density in this image  are 
$25 \,{\rm mJy/Beam}$ and $-45 \, {\rm mJy/Beam}$ respectively.  
The subsequent analysis was done using the visibility data. We have 
analyzed the data both before and after the sources were subtracted,
and we shall refer to these as {\bf data I} (Initial - before
source subtraction)  and {\bf data R} (residual - after source
subtraction) respectively.    

The final data   contains $295868$ baselines, each of which has 
visibilities for  $2$ circular polarizations and $96$ frequency 
channels, of which we have used  only the first $80$  channels 
for the subsequent analysis.  The visibilities from the two 
polarizations were combined for the subsequent analysis.  
The  real and imaginary parts of the resulting visibilities 
have a mean value $-0.56 \, {\rm mJy}$ and $2.6 \, {\rm mJy}$ 
respectively ,   and  rms of $2.93 \, {\rm Jy}$ for both 
in  {\bf data I}. For {\bf  data R} the real  and imaginary 
parts  of the visibilities have a mean   $-6.0 \, {\rm   mJy}$ 
and   $1.1 \, {\rm mJy}$  respectively whereas the  
rms is $2.42 \, {\rm Jy}$ for both. 

\begin{figure*}
\includegraphics[width=75mm]{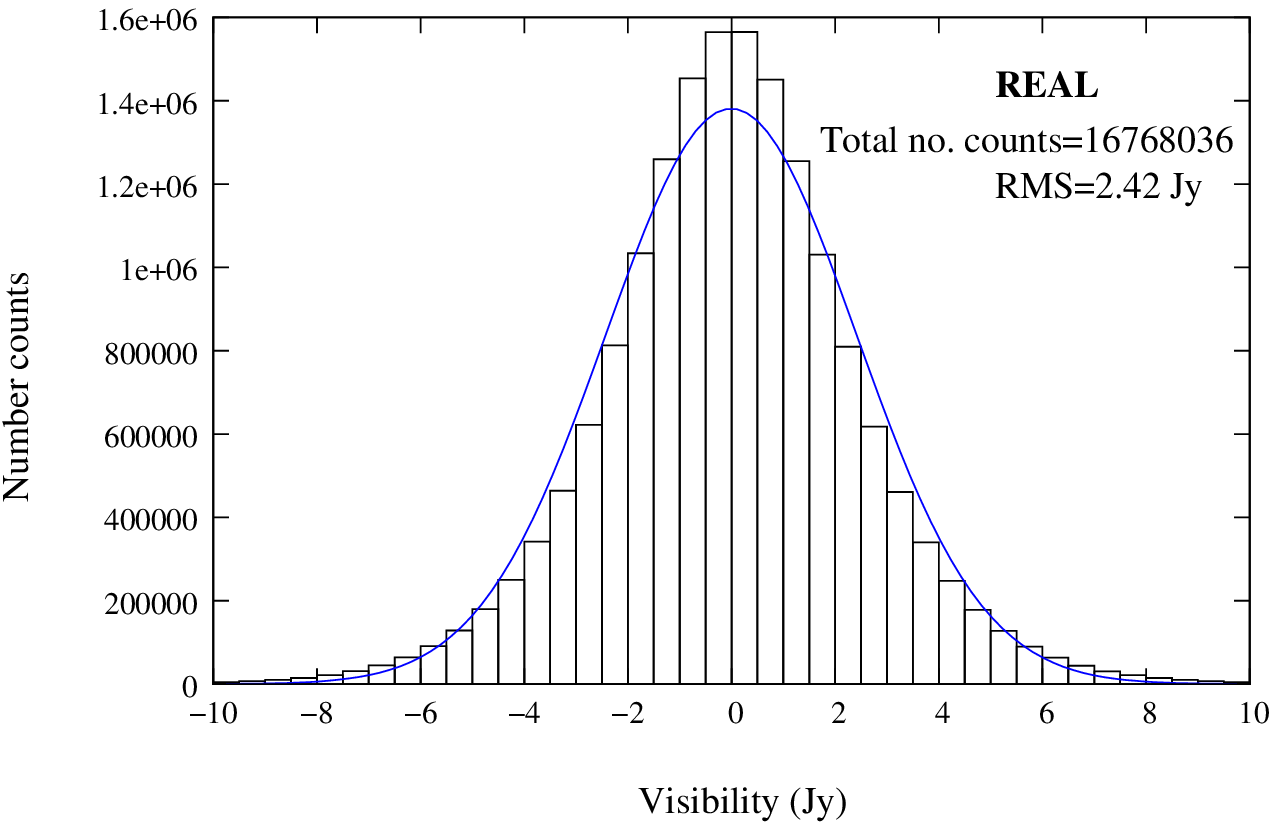}
\includegraphics[width=75mm]{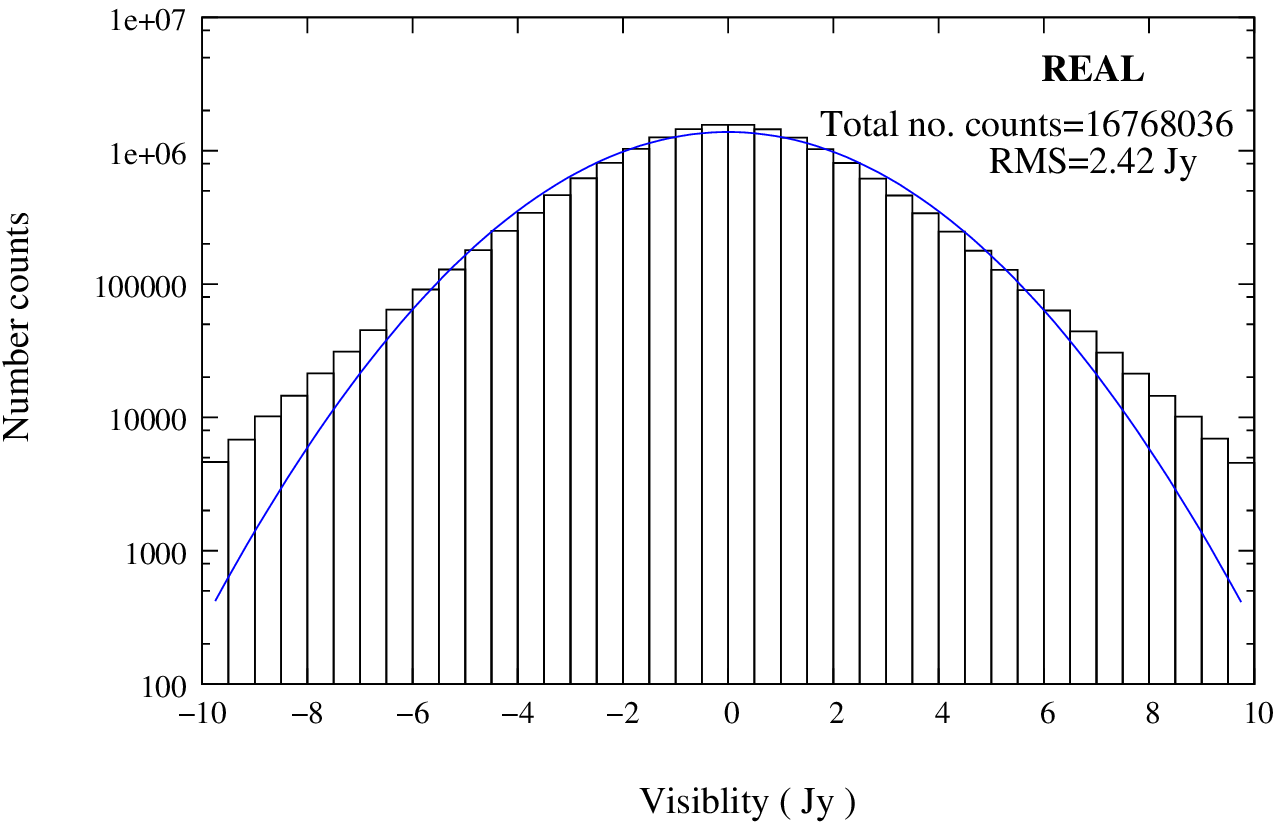}
\caption{The distribution of  visibilities after
  source subtraction ({\bf{data R}}). The same plot is shown on a
  linear scale (left panel) and a log-linear scale (right panel). 
The data is plotted as a histogram, and a Gaussian with the
  corresponding mean and rms. (see text) is plotted as a solid line.  
The discrepancy at high amplitudes ($\ge 6$ {\rm Jy}) is visible only
  in the right panel.} 
\label{fig:h1}
\end{figure*}

In the subsequent analysis it is often convenient to assume that the
visibilities  have a Gaussian distribution. Figure \ref{fig:h1} shows
the distribution of the real part of the visibilities for {\bf data
  R}. We find that a Gaussian gives a reasonably good fit to the data
within $2 \sigma$ which contains the bulk of the data. 
 The number counts predicted by the Gaussian falls
much faster than the data at large visibility values $\mid Re(V) \mid  >
6 \, {\rm Jy}$. Deviation from Gaussian statistics  is expected to
mainly affect the error estimate on the visibility correlation. We
expect this effect
to be small, since only a small fraction of visibilities are discrepant.
The imaginary part of of {\bf data R}, and the real
and imaginary parts of {\bf data I} all show a similar behaviour.

\section{ Visibility correlations}

The visibility $V(\u,\nu)$ measured in a radio-interferometric  
observation is the sum of three different contributions   
\begin{equation}
V(\u,\nu)=S(\u,\nu)+ F(\u,\nu) + N(\u,\nu)
\end{equation}
the HI signal $S(\u,\nu)$, astrophysical foregrounds $F(\u,\nu)$ and
system noise $N(\u,\nu)$. We treat all three of these contributions as 
uncorrelated random variables with zero mean. The statistical
properties of the visibility can be quantified through the two
visibility correlation (henceforth the visibility correlation)

\begin{equation}
V_2(\u_1,\nu_1;\u_2,\nu_2)=\langle V(\u_1,\nu_1) V^{*}(\u_2,\nu_2)
\rangle 
\end{equation}
and 
\begin{equation}
V_2=S_2+F_2+N_2
\end{equation}

where $S_2$, $F_2$ and $N_2$ respectively refer to the signal,
foreground and noise contributions to the visibility correlation. 

The contribution from the HI signal $S_2$ is expected to be 
$\sim 10^{-7} \, {\rm Jy}^2$ or smaller at $150 \, {\rm MHz}$
\citep{BA5}. This is negligible compared to the expected foregrounds 
and noise contributions in our observations, and hence we 
ignore it in our further analysis.

The foreground contribution $F(\u,\nu)$ is the Fourier transform of
the product of the foreground specific intensity distribution on the
sky $ I(\th,\nu)$ and the primary beam pattern of the individual GMRT
antenna $A(\th,\nu)$. As mentioned earlier, this Fourier relation is
strictly valid only if the field of view is small, and in this
observation we expect  considerable deviations at large baselines.  
As we are mainly interested in the visibility correlations at small
baselines, and also because the analysis is considerably more
complicated otherwise, we assume the Fourier relation to hold. We can
then express $F(\u,\nu)$ as a convolution
\begin{equation}
F(\u,\nu)= \int \tilde{I}(\u^\prime,\nu)\,\tilde{a}(\u -\u^\prime,
\nu)\, d^2 \u^\prime. 
\label{eq:a2}
\end{equation} 
 where $\tilde{I}(\u ,\nu)$ and $\tilde{a}(\u , \nu)$
 are  the Fourier transform of $I(\th,\nu) $ and $A(\th,\nu)$
 respectively. Assuming that the region of sky under  observation is
 small so that it can be treated as flat, we have  
\begin{eqnarray}
\langle \tilde{I}(\u_1,\nu_1)  \tilde{I}(\u_2,\nu_2)\rangle &=& \delta^2_D(\u_1
 -\u_2) \, 
\left(\frac{\del B}{\del
  T}\right)_{\nu_1} \left(\frac{\del B}{\del T}\right)_{\nu_2} 
\, \nonumber \\  &\times  
  & C_{2 \pi U_1} (\nu_1,\nu_2)
\end{eqnarray}
where $\delta^2_D(\u_1  -\u_2)$ is the  two dimensional Dirac Delta
 function, $(\del B/\del  T)_{\nu}=2 k_B \nu^2/c^2$ is the
conversion factor from brightness temperature to specific intensity 
 and  $C_l(\nu_1,\nu_2)$ is the multi-frequency angular power
 spectrum (MAPS; eg. \citealt{kanan})
of the foreground brightness temperature distribution.
 Using this to calculate the 
 foreground contribution to the visibility correlation we have 
\begin{eqnarray}
F_2(\u_1,\nu_1;\u_2,\nu_2) &= & \int
  d^2 U^\prime\,\tilde{a}(\u_1-\u^\prime ,\nu_1)\,
  \tilde{a}^*(\u_2-\u^\prime ,\nu_2)\, \nonumber \\  &\times  
  & \left(\frac{\del B}{\del
  T}\right)_{\nu_1} \left(\frac{\del B}{\del T}\right)_{\nu_2} 
C_{2 \pi U^\prime }(\nu_1,\,\nu_2) \,.
\label{eq:a4}
\end{eqnarray} 
The GMRT primary beam is well parametrized by a Gaussian
 $A(\th,\nu)=e^{-\theta^2/\theta_0^2}$ where $\theta_0 \approx
0.6 \times  \theta_{\rm FWHM}=2.3^{\circ}$.  There is  a small
 variation in  $\theta_0$   ($ \propto \nu^{-1}$) across the
 frequency  band.  Ignoring  this  $\nu$ dependence have 
 $\tilde{a}(\u,\nu) =  \tilde{a}(\u)=\pi \theta_0^2
 \exp[-\theta_0^2\pi^2 U^2]$. 
The integral in   eq. (\ref{eq:a4})   has a very small value unless
 the terms  $  \tilde{a}(\u_1-\u^\prime )\,$ and 
$ \tilde{a}^*(\u_2-\u^\prime )  $  have a considerable overlap 
{\it  ie. } $\mid \u_1 -\u_2 \mid < (\pi \theta_0)^{-1}$. This tells us
 that $F_2(\u,\nu_1;\u+\Delta \u,\nu_2)$ has a significant value only
 if $|\Delta \u| < (\pi \theta_0)^{-1}$ and is negligible
 otherwise. Further,  $\mid \Delta \u \mid  \ll U$ at the
 baselines of interest, and we may approximate $a^*(\u+\Delta \u
 -\u^\prime) \approx a^*(\u  -\u^\prime)$ in eq. (\ref{eq:a4}) and
 write 
\begin{eqnarray}
F_2(\u,\nu;\u + \Delta \u,\nu+\Delta \nu) &= &  \left(\frac{\del
  B}{\del   T}\right)^2_{\nu}  
\int
  d^2 U^\prime\, |\tilde{a}(\u-\u^\prime )|^2\, \nonumber \\  &\times  
  &
C_{2 \pi U^\prime }(\nu,\,\nu+\Delta \nu) \,.
\label{eq:a5}
\end{eqnarray} 
where we have ignored the $\Delta \nu$ dependence of $\theta_0$ and $\left(
  \frac{\del    B}{\del   T}\right) $.   

The explicit reference to $\Delta \u$ can be dropped as it does not
appear in the integral. 
We also assume that $C_{2 \pi U}(\nu_1,\nu_2)$ is a slowly varying
function of $U$ as compared to $|\tilde{a}(\u)|^2$ whereby  
$|\tilde{a}(\u-\u^\prime )|^2 \approx (\pi \theta_0^2/2)
\delta^2_D(\u-\u^\prime ) $ which gives   
\begin{equation}
F_2(\u,\Delta \nu)=\frac{\pi\theta_0^2}{2}
 \left(\frac{\del B}{\del   T}\right)^2
C_{2\pi U}(\Delta \nu) \,Q(\Delta \nu) 
\label{eq:a8}
\end{equation}
where $Q(\Delta \nu)$ incorporates the effect of the $\Delta \nu$
dependence of $\theta_0$ and$\left(  \frac{\del    B}{\del
  T}\right) $.   
We are mainly interested in the $\Delta \nu$ dependence,  and
we do not show the $\nu$ dependence explicitly. 
Equation  (\ref{eq:a8})  relates the angular power spectrum of the
foreground contribution to the visibility correlations which can be
determined from our observations. 

The system noise makes a contribution 
\begin{equation}
N_2(\u_1,\nu_1;\u_2,\nu_2)=\delta_{\u_1,\u_2} \delta_{\nu_1,\nu_2}
\langle N^2 \rangle 
\label{eq:n2}
\end{equation}
which is non-zero only when a particular visibility  is correlated 
with itself.  For a single polarization, the 
rms. noise in   the real part (or equivalently  the imaginary part) of
a visibility  is  expected to be \citep{thompson}
\begin{equation}
\sigma
=\frac{\sqrt2k_BT_{sys}}{A_{eff}\sqrt{\Delta \nu
      \Delta t}}
\label{eq:rms}
\end{equation}
where $T_{sys}$ is the total system temperature, $k_B$ is the Boltzmann
constant, $A_{eff}$ is the effective collecting area of each
antenna, $\Delta \nu$ is the channel width and $\Delta t$ is correlator
integration time. For the GMRT
parameters\footnote{http://www.gmrt.ncra.tifr.res.in} 
 this is predicted to be 
$\sigma=1.03 \, {\rm Jy}$ for a single polarization. We have combined
 both polarizations, and so the variance in each  visibility of the
 final data that we have analyzed is $2 \, \sigma^2$. In
 eq. (\ref{eq:n2}) the variance of the 
 real  and  imaginary parts of the noise in a  visibility  contribute
 in quadrature and  we have $\langle N^2 \rangle = 4 \sigma^2$.

\subsection{Estimating the visibility correlation.}
We use the estimator
\begin{equation}
\hat{V}_2(U,\Delta \nu)=\overline{V(\u,\nu_i) V^{*}(\u+\Delta
  \u,\nu_i+\Delta   \nu)}
\end{equation}
where the bar  denotes an average over the data under the
assumptions 
\begin{enumerate}
\item The $\u$ dependence is isotropic {\it ie.} $V_2$ depends
only on the magnitude $U$ and not  the direction of $\u$ 
\item The $\Delta \nu$ dependence is the same if the frequency origin
  $\nu_i$ is shifted to another channel $\nu_j$ in the observation
  frequency band. 
\item Only visibilities $V(\u+\Delta \u,\nu_i + \Delta \nu)$ at
  baselines $\u+\Delta \u$ within a disk of radius $\mid 
  \Delta \u \mid \le 
  D < (\pi   \theta_0)^{-1}$ centered at  $\u$ are correlated with
  $V(\u,\nu_i)$, and   $\hat{V}_2$  is averaged over   this disk.  
\end{enumerate}
Note that the second assumption above implies that $\hat{V}_2(U,\Delta
 \nu)$ gives  an  estimate  of the average
$\Delta \nu$ dependence across the entire frequency band. It also
 implies an average  over positive and negative $\Delta \nu$ values.  
Besides this, the estimator is averaged over bins in $U$
$(U_1-U_2, U_2 -U_3,...)$. so that we have
$\hat{V_2}(U_i,\Delta \nu)$ at a few values $U_i$ corresponding to the
average baseline of the bins. 

The correlation of a visibility with itself introduces a noise
contribution in the expectation value of this estimator. The noise
contribution can be avoided (eg. \citealt{Begum}) by excluding
self-correlations {\it ie.} the visibility
$V(\u,\nu_i)$ is correlated with every baseline $V(\u+\Delta
\u,\nu_i+\Delta \nu)$ within  a disk $|\Delta \u| <D$ except itself. 
The expectation value of the estimator has a value  
\begin{equation}
\langle \hat{V}_2(U,\Delta \nu)\rangle = F_2(U,\Delta \nu) \,.
\end{equation}
which provides an unbiased estimate of the foregrounds. 
 The system noise makes a contribution  only to the uncertainty or the
 error in the  estimator. The expectation value of the estimator is
 real. The value of the estimator determined from an observation will,
 in general,  have a real and an imaginary part. The real part
 contains the foreground information, whereas the imaginary part 
of the observed value of the estimator
 can be attributed to statistical fluctuations in the foregrounds and
 the noise. 
 
\subsection{Error Estimates}
The expected uncertainty or statistical fluctuations in the real part
of the estimator 

\begin{equation}
  \sqrt{\langle (\Delta \hat{V_2})^2 \rangle} \equiv \sqrt{\langle
    (\hat{V_2}-\langle \hat{V_2}\rangle )^2 \rangle} 
\end{equation}

is the  sum of two contributions 

\begin{equation}
\langle (\Delta \hat{V_2})^2 \rangle = (\Delta F_2)^2  + (\Delta N_2)^2 \,.
\end{equation}
If we assume that the foregrounds are a Gaussian random field, the
foreground contribution to the error is 
\begin{equation}
[\Delta F_2(U_i,\Delta \nu)]^2=\frac{1}{N_E}  \left[
\frac{F_2(U_i,0) + F_2(U_i,\Delta \nu)}{2} \right]
\label{eq:b4}
\end{equation}
where $N_E$ is the number of independent estimates of $F_2(U,\Delta
\nu)$ that  contribute to $ \hat{V}_2(U_i,\Delta \nu)$.   The
baselines within  a disk of radius $\sim (\pi \theta_0)^{-1}$ in $uv$
space (Figure \ref{fig:a1}) are correlated, and all the baselines
within such a disk provide only one independent estimate of the
visibility correlation. For each $U$  bin $N_E$ is determined by
counting  the number of such  regions  with the $uv$ coverage of our
observations. 

The system noise contribution in any two visibilities are
uncorrelated, and hence 
\begin{equation}
[\Delta N_2(U,\Delta \nu)]^2 
=\frac{\langle N^2\rangle^2}{2 N_P}=\frac{8 \sigma^4}{N_P}
\label{eq:b5}
 \end{equation}
where $N_P$ is the number of visibility pairs  that contribute to 
the estimator $\hat{V}_2(U,\Delta \nu)$ for a particular $U$ bin and
$\Delta \nu$ separation. 

The error  in the imaginary part of the estimator  also is a sum
of two contributions. The foreground contribution is somewhat
different from   eq. (\ref{eq:b4}) and we have 
\begin{equation}
[\Delta F_2(U_i,\Delta \nu)]^2=\frac{1}{N_E}  \left[
\frac{F_2(U_i,0) - F_2(U_i,\Delta \nu)}{2} \right]
\end{equation}
while the system noise contribution is the same a eq. (\ref{eq:b5}).

\section{Foreground model predictions}

We consider only the two most dominant  foreground
components namely extragalactic radio sources and the 
diffuse synchrotron radiation from our own Galaxy. The 
free-free   emissions from our Galaxy and external galaxies is around 
 $1\%$ of the total foreground contribution \citep{shaver}, and we
ignore this in our analysis. 
For each foreground component the MAPS can be  modeled as 
\begin{equation}
C_{l}(\nu_1, \nu_2)=A \left(\frac{\nu_f}{\nu_1} \right)^{
  \bar{\alpha}} 
\left(\frac{\nu_f}{\nu_2} \right)^{ \bar{\alpha}}
\left(\frac{1000}{l}\right)^{\beta}  I_l(\nu_1\,, \nu_2)
\label{eq:fg}
\end{equation}
where  $\nu_f=130 \,{\rm  MHz}$,  and for each foreground
 component 
 $A$, $\beta$ and $\bar{\alpha}$
are the amplitude, the power law index of the angular power spectrum 
 and the mean spectral  index respectively.  The actual spectral  
index varies with line of sight across the sky and this causes the
foreground contribution to decorrelate with increasing frequency
separation $\Delta \nu=\mid \nu_2 - \nu_1 \mid$ which is quantified
 through the 
foreground frequency  decorrelation function $I_l(\nu_1\, \nu_2)$
\citep{zald}  which has been  modeled as 
\begin{equation}
I_l(\nu_1\,, \nu_2)=\exp\left[ - \log_{10}^2 \left(\frac{\nu_2}{\nu_1}
 \right)/2   \xi^2 \right] \,.
\end{equation}
The model parameters values that we have
    used are discussed below and are given  in Table~\ref{tab:parm}. 

Resolved extragalactic radio sources (point sources) dominate the
 radio sky at $150 
 \, {\rm MHz}$. \citet{dmat1} have used the 6C survey
 \citep{hales}, and the 3CR survey  and the  3 CRR
catalogue  \citep{laing}  to estimate this contribution. The
 limiting flux density of these surveys  was
 $\sim 100\, {\rm  mJy}$ and the extrapolation to fainter sources is
 rather  uncertain. \citet{dmat1} have fitted the differential source
 counts using   a double power-law with the change in slope  occurring
 at $ 880 \, {\rm mJy}$. Since the brightest source in our image  has a flux
 density below $880 \, {\rm mJy}$ we use only the fit to the fainter
 part
\begin{equation}
  \frac{dN}{dS}= \frac{4000}{Jy \cdot Sr}\cdot\,\left(\frac{S}{1
  Jy}\right)^{-1.75}\,. 
\label{eq:a11}
\end{equation} 
These sources make two distinct contributions to MAPS, the first being 
 the Poisson noise arising from the discrete nature of these sources
 and the second arising from the  clustering of the
 sources. Table~\ref{tab:parm} shows  the respective parameters based
 on  the estimates of 
 \citet{dmat1} who assume that these sources are clustered like
  galaxies today or as Lyman-break galaxies (\citealt{giava}) at $z
 \sim 3$.  Using these in eq. (\ref{eq:a8}) to calculate the
 foreground contribution to the visibility correlation at $153 \,{\rm
 MHz}$  for $\Delta \nu=0$, we have the Poisson term  
\begin{equation}
 F_2(U,0) =7.6\,\,\, \left(\frac{s_c}{J_y}\right)^{1.25}
 \,\,\,\,{\rm Jy}^2 \,,
\label{eq:12a}
\end{equation}
and the clustering term 
\begin{equation}
 F_2(U,0)  =0.51\,\,\left(
 \frac{S_c}{J_y}\right)^{0.5}\,\left(\frac{U}{1000}\right)^{-1.1} \,\,\,\,\,{\rm
 J_y}^2  \,.    
\label{eq:12b}
\end{equation}
Here it is assumed that sources with flux  greater than $S_c$
 have been identified from continuum images and removed   from the  
 data. The brightest source in our initial image has  $S \sim 890 \, {\rm
 mJy}$ and we use this value for $S_c$ when comparing model
 predictions with results from {\bf data I}. For {\bf data R} we have
 used $S_c=8\, {\rm mJy}$ as we have used this as the limiting value
 for our source subtraction (Section 2). 

The uncertainty or error in the model prediction for these radio
sources is also a sum of two parts. The error in the clustering part
can be estimated using eq. (\ref{eq:b4}). For the Poisson part the
variance of $F_2$ involves  the fourth moment of the differential
source count and we have 
\begin{equation}
[\Delta F_2(U,0)]^2=\left(\frac{s_c}{{\rm Jy}}\right)^{2.5}\left[\,\, 63.2-
 1.54\,\left(\frac{s_c}{{\rm Jy}}\right)^{0.75}\,\, \right] . 
\end{equation}
\vspace{.2in}

\begin{table}
\caption{Fiducial values of the parameters  used for characterizing
  different foreground contributions} 
\vspace{.2in}
\label{tab:parm}
\begin{tabular}{|c|c|c|c|c|}
\hline
Foregrounds & $A ({\rm mK^2})$ & $\bar{\alpha}$ & $\beta$ & $\xi$\\ 
\hline
Point source  & $1.2 \times 10^4\left (\frac{S_{cut}}{\rm
    Jy}\right )^{1.25}$ & $2.07$ & $0$ & $1$ \\
(Poisson part) & & & & \\
\hline
Point source & $6.1 \times 10^3\left (\frac{S_{cut}}{\rm Jy}\right
)^{0.5}$  & $2.07$ & $1.1$ &$2$ \\
(clustered part) & & & & \\
\hline
Galactic synchrotron & $700$ & $2.80$ &$2.4$ & $4$ \\
\hline
\end{tabular}
\end{table}
\vspace{.2in}

 The diffuse Galactic synchrotron radiation 
is believed to be produced by cosmic ray electrons propagating
in the magnetic field of the Galaxy \citep{GS69}.This  has an angular
power spectrum that scales as $C_l \sim l^{-2.4}$ \citep{teg00},
 though this slope ($\beta$) is rather uncertain. 
The   analysis of  radio surveys at 408 MHz, 1.42 GHz, and 2.326
GHz \citep*{haslam,R82,RR88,JBN98} show the 
spectral index to be $\bar{\alpha} \approx 2.8$  which is in general
agreement with result of \citet{platania1}. For the synchrotron
radiation,  in Table~\ref{tab:parm} we
have adopted the parameters from \citet{santos} which gives 
\begin{equation}
 F_2(U,0) =4.2\times10^{-3}\,\left(\frac{U}{1000}\right)
^{-2.4}\,\,\,\,\,{\rm Jy}^2.  
\label{eq:a14}
\end{equation}
We note that the amplitude of the synchrotron contribution is very
sensitive to the spectral index whose value is quite uncertain.
The  value  is in the range  $2.5 \le \bar{\alpha}  \le 3$,
and the amplitude increases by nearly an order of magnitude if
$\bar{\alpha}=3$ instead of $\bar{\alpha}=2.8$ as assumed here.  

The error for the synchrotron prediction can be calculated using
eq. \ref{eq:b4}. The total error in the model predictions is
calculated by adding the variances from the different contributions. 

For the frequency separations of our interest $(\Delta \nu < 2.5 \,
{\rm MHz})$,  for all the foreground components 
 the $(\nu_2/\nu_f)^{\bar{\alpha}}$ term in equation
(\ref{eq:fg}) introduces a larger $\Delta \nu$ dependence  in
 $C_l(\Delta \nu)$ as compared to the frequency decorrelation function
 $I(\nu_1,\nu_2)$. When calculating $F_2(U,\Delta \nu)$ 
it is necessary to also  incorporate  $Q(\Delta \nu)$ 
(eq. \ref{eq:a8})  which has the $\Delta \nu$ dependence arising from
$\theta_0$ and  $(\del B/\del  T)_{\nu}$.  All
of these predict a smooth $\Delta \nu$ dependence, and we may use a
Taylor series expansion 
\begin{equation}
F_2(U,\Delta \nu)=F_2(U,0)\, \left[1 \, + \,
B \left(\frac{\Delta \nu}{\nu}\right)^2 ...\right]
\label{eq:m1}
\end{equation}
where $B$ is a constant of order unity. The 
$\Delta    \nu/\nu$ term does not appear in eq. (\ref{eq:m1}). 
This term cancels out  because the estimator $\hat{V}_2(U,\Delta \nu)$
averages positive and negative $\Delta \nu$ values.  
We use  $B=1$ to make an order of magnitude estimate. The
expected change in $F_2(U,\Delta \nu)$  is $\sim 3 \times
10^{-2} \, \%$ for  $\Delta \nu=2.5 \, {\rm MHz}$.  The
key point here is that $F_2(U,\Delta \nu)$ is predicted to change very
slowly with $\Delta \nu$, and the change is also very small.

\section{Results and Discussion}
\begin{figure*}
\includegraphics[width=70mm,angle=-90]{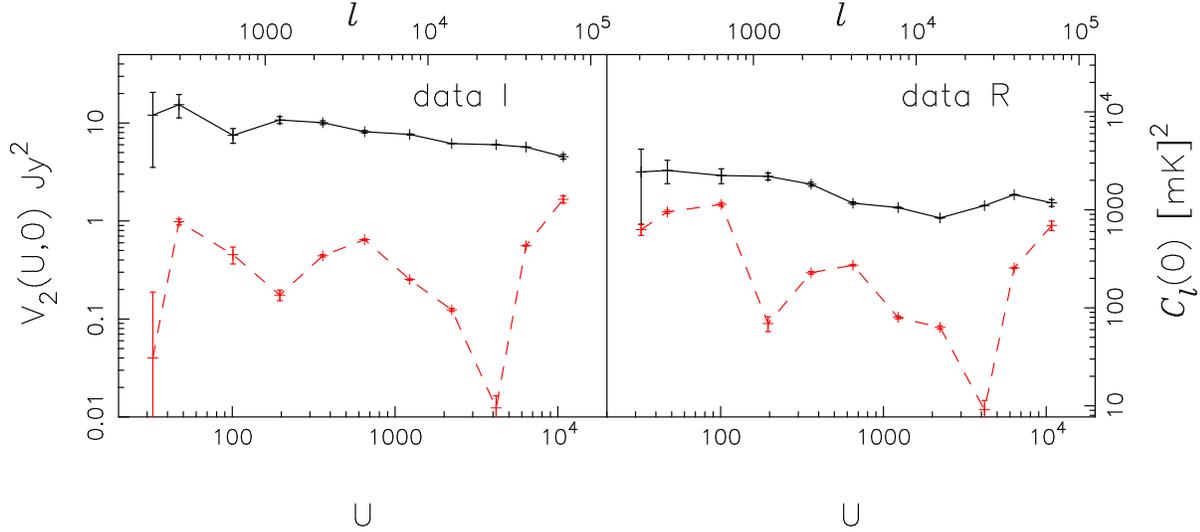}
\caption{This shows the real (upper curve) and imaginary (lower curve)
  parts of the observed visibility correlation $V_2(U,0)$ as a
  function of $U$ for the two   data-sets indicated in the figure. 
As shown here, this may also be interpreted as $C_l(0)$ as a function
  of $l$.}  
\label{fig:r1} 
\end{figure*}

We have determined the observed value  $V_2(U,\Delta \nu)$ 
of the visibility correlation  estimator $\hat{V}_2(U,\Delta \nu)$
for    {\bf data I} and {\bf data  R}  which are before
and after source subtraction  respectively. Baselines
in the range $20 \le U \le 2 \times 10^4$, and frequency channels $21$ 
to $100$ were used for the analysis. 
Visibilities  $V(\u+\Delta
\u,\nu + \Delta \nu)$ within the  disk  $\mid \Delta \u \mid \le
D=5$ were correlated with $V(\u,\nu)$. Here  $\Delta \nu$ was
restricted to $\mid \Delta \nu \mid < 2.5 \, {\rm MHz}$ which
corresponds to  a separation of $40$ channels. 
Note that the correlation of a visibility with itself was not included.   
The value of $D$ was chosen such that it is both less than $(\pi \theta_0)^{-1}=8$,
and  also large enough that a reasonable number of visibility pairs
that contribute to the correlation. Figure \ref{fig:r1} shows $V_2(U,\Delta \nu)$
as a function of $U$ for  $\Delta \nu=0$.  Equivalently, we may
also interpret this as   the multi-frequency angular  power spectrum
$C_l(\Delta \nu)$  at $\Delta \nu=0$. 

 For both the data-sets  the real part of $V_2(U,0)$
is found to be considerably larger than the imaginary part. 
This  is consistent with  the discussion of  Section 3.1, and we
expect the real  part to provide an estimate of the foreground
contribution $V_2(U,0)$.  The $1-\sigma$ error bars shown in the
figure have been determined based on the error estimates discussed in
Section 3.2. The uncertainty in $F_2(U,0)$ is mainly  due to  the
limited number of independent estimates, the system noise
makes a smaller contribution.  Though the results for {\bf data I}
over the range $200 \le U \le \le 2 \times 10^4$
looks like a power law $V_2(U,0)\propto U^{-\alpha}$ with a very small
slope $0 \le \alpha \le 0.25$,  we do not find a fit with an acceptable
value of $\chi^2$ per degree of freedom. 

The real part of $V_2(U,0)$ falls to nearly one-fourth of its original
value at most of the  $U$ bins when the directly detected sources are
subtracted 
out. This indicates that a large part of the contribution to 
$V_2(U,0)$ in {\bf data I}  is from these resolved sources, and we
may interpret $V_2(U,0)$ as arising primarily from  these
sources. {\bf Data R} is expected to contain contributions from point 
sources below  the detection limit of our image, diffuse sources, 
system noise, limitations in our imaging and source subtraction 
procedure and residual RFI. We will assume for the moment that
these effects can be ignored, but return to this issue later
in this section. 

\begin{figure*}
\includegraphics[width=70mm,angle=-90]{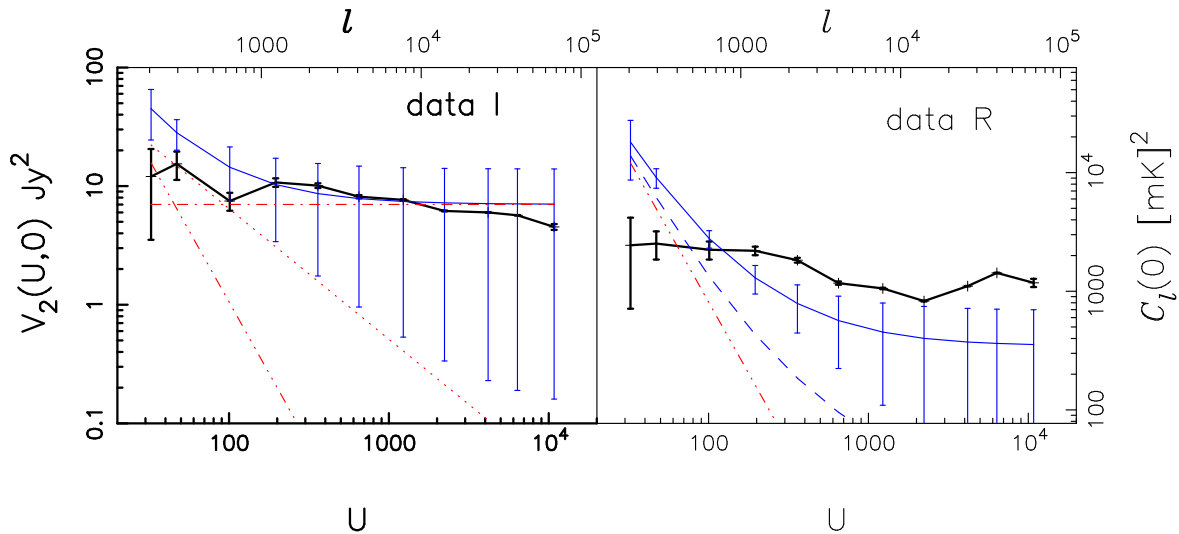}
\caption{The thick solid line shows the real  part of the
  observed  
  visibility correlation $V_2(U,0)$ as a function of $U$ for the
  two data-sets indicated in the figure. As shown here, this may
  also be interpreted as   $C_l(0)$ as a function of $l$.  For {\bf
  data I} the thin solid line  shows the total model prediction 
  for $S_c=900  \, {\rm mJy}$. Also shown are  the  
  contributions from 
point source Poisson (dash-dot), point source clustering (dot) and
  Galactic synchrotron (dash-dot-dot-dot).  For {\bf data
  R} the  thin solid line shows the total model predictions for
  $S_c=100 \, {\rm mJy}$     and and the long dashed line for   
$10 \, {\rm mJy}$. The dash-dot-dot-dot curve shows the 
  Galactic synchrotron contribution.}  
\label{fig:r2}	  
\end{figure*}

Figure 	\ref{fig:r2} shows the observed $V_2(U,0)$ plotted against the
predictions of the foreground models discussed in Section 4.  The
brightest source in our image has  flux $890 \, {\rm mJy}$. Based on
this we use  $S_c=900 \, {\rm mJy}$ for the point  source contribution
to {\bf data I}.  The clustering of point sources  dominates
at baselines $U<150$ ($\theta >0.7^{\circ}$), while the 
Poisson fluctuations  of the point sources dominates at larger 
baselines.  The diffuse Galactic synchrotron radiation is much smaller
than the point source contribution  at all baselines. The errors in
the model prediction are quite large and are mainly  due to the
Poisson fluctuations of the point sources. The model predictions are
found to be  consistent with  the observed values of
$V_2(U,0)$ except at the smallest $U$ value which corresponds to an 
angular scale of $\sim 1.8^{\circ}$.  At these baselines 
the convolution with the primary beam
pattern (eq. (\ref{eq:a4})) becomes important. We have not included
this, and the  actual model predictions would  possibly be somewhat
smaller if  this were  included. 
As noted in Section 4., the amplitude of the synchrotron
contribution  is very sensitive to the value of the spectral
index. The amplitude decreases by a factor of $\sim 18$ if
$\bar{\alpha}=2.5$ instead of the value $\bar{\alpha}=2.8$ used here. 
This changes the total foreground contribution only at small baselines
($U \le 100$) where the model then becomes consistent with our
observations. 

The limiting flux for source subtraction is $\sim 8 \, {\rm mJy}$, and 
hence we use  $S_c= 10 \, {\rm mJy}$  for
{\bf data R}. The model prediction is dominated by 
Galactic synchrotron radiation at $U < 150$,  point source clustering in the
range $150 \le U \le 2 \times 10^3$ and point source Poisson
fluctuations at $U > 2 \times 10^3$.
 The model predictions fall short of the observations at all
 baselines except the smallest $U$ value where it overshoots the
 observations.  Since the model prediction for $S_c=10 \, {\rm mJy}$
 falls very much short of  the observations, we also consider 
$S_c=100 \, {\rm mJy}$ where the
dominant contribution is 
Galactic synchrotron at $U < 60$,  point source clustering in the
range $60 \le U \le 400$ and point source Poisson
fluctuations at $U > 400$. We find that the observations are slightly
above the $1-\sigma$ error-bars at  baselines  $U > 100$,
whereas they exceed the model predictions at baselines $U<100$. 
A point to  note is that at the smallest
baseline the  prediction for the Galactic synchrotron  radiation
exceeds the  observation. This may  be a consequence of the
possibility  that the background radiation is relatively  low in 
the direction of our observation. Estimates from the  \citet{haslam}
map at $408 \, {\rm MHZ}$ show a relatively low brightness temperature
of $\sim 30 \, {\rm K}$ towards the direction of our observation. 
 
We quantify the $\Delta \nu$ dependence of $V_2(U,\Delta \nu)$
 using  $\kappa(U,\Delta \nu)$ which is defined as 
\begin{equation}
\kappa(U,\Delta \nu)=\frac{V_2(U,\Delta \nu)}{V_2(U,0)}\,.
\end{equation}

\begin{figure*}
  \includegraphics[width=100mm,angle=-90]{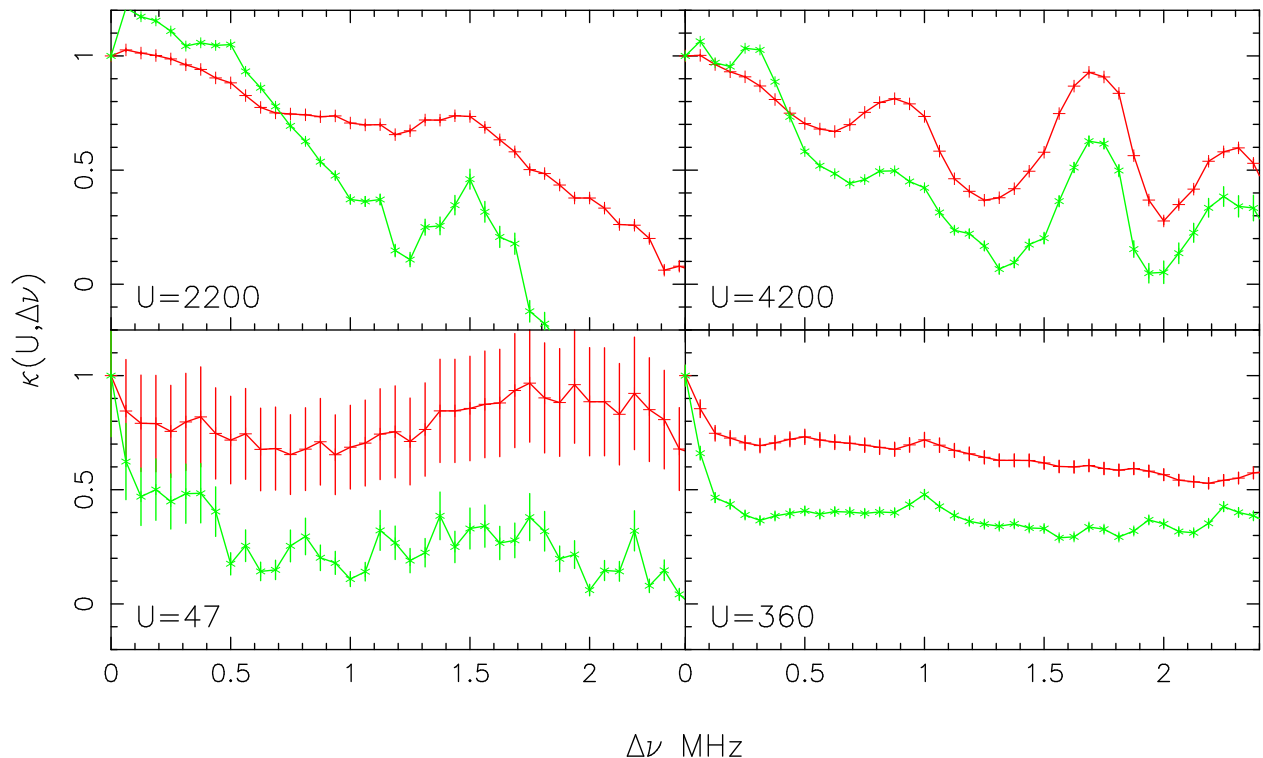}
  \caption{This shows $\kappa(U,\Delta \nu)$ as a function of $\Delta
  \nu$ for the different $U$ values shown in the figure. The upper
  curve (at large $\Delta \nu$)  shows {\bf data I} while the lower
  shows {\bf data R}.} 
  \label{fig:r3}
\end{figure*}

We expect the visibilities $V(\u,\nu)$  and $V(\u,\nu+\Delta \nu)$ to
get decorrelated as $\Delta \nu$ is increased, and hence we expect $0
\le \mid \kappa(U,\Delta \nu) \mid \le 1$.  
Figure \ref{fig:r3} shows $\kappa(U,\Delta \nu)$ for different values
of $U$. The foreground models predict a smooth $\Delta \nu$
dependence for $\kappa(U,\Delta \nu)$.  The departure from 
$\kappa(U,\Delta  \nu)=1$ is predicted to be less than $1 \%$  for
$\Delta \nu < \, 2.5 \,{\rm MHz}$.  The observed behavior of
$\kappa(U,\Delta \nu)$ is quite different from the model
predictions.  At the small baselines $U < 1000$ we find that
$\kappa(U,\Delta \nu)$ falls sharply within  the first three  channels.
In the $U=47$ bin $\kappa(U,\Delta \nu)$ fluctuates at large $\Delta
\nu$ whereas it  
remains roughly constant at $U=360$. In both cases 
this  value of
$\kappa(U,\Delta \nu)$  is smaller for {\bf  data R} as  compared to {\bf
  data I}.  At $U=2200$, for {\bf data I}  
$\kappa(U,\Delta \nu)$  falls gradually  with increasing $\Delta \nu$,
and the  visibilities are uncorrelated ($\kappa(U,\Delta \nu)\sim 0$)
by $\Delta \nu \sim 2.5 \, {\rm MHz}$. Interestingly, for {\bf data R}
we find that $\kappa(U,\Delta \nu)$ shows a sudden increase to 
$\kappa(U,\Delta \nu)>1$ at very small $\Delta \nu$ ($< 0.5 \, {\rm
  MHz}$), after which  $\kappa(U,\Delta \nu)$ falls and becomes
negative by $\Delta \nu \sim < 2 \, {\rm  MHz}$. 
It appears that in this $U$ bin our source subtraction procedure
has introduced excess correlations between the visibilities at 
small $\Delta \nu$ and introduces anti correlations at large 
$\Delta \nu$. At $U=4200$, for {\bf data I}  the
value of $\kappa(U,\Delta \nu)$ oscillates with increasing $\Delta
\nu$. At large $\Delta \nu$ {\bf data  R} also shows  a similar
behavior except that the  $\kappa(U,\Delta \nu)$ values are smaller. 
The behavior of {\bf data R} is quite different
from that of {\bf data I} at  very small $\Delta \nu$  where there are 
two small oscillations that cross $\kappa(U,\Delta \nu)=1 $. 

The first point that emerges from our results is that the observed
visibility correlations $V_2(U, 0)$ is consistent with the predictions
of the  existing foreground models at all baselines  
except the smallest one  which probes angular scales $\sim
1^{\circ}$. The observations are in excess of the model prediction at
the smallest baseline. 
 The second point is that $V_2(U,\Delta \nu)$ 
shows considerable $\Delta \nu$ dependence,  there being changes of
order unity within $\Delta \nu= 2.5 \, {\rm MHz}$.  This rapid change in
the visibilities $V_2(U,\nu)$ across frequency channels is contrary to
the foreground models which predict changes less than $1 \%$.

It is well appreciated that accurate subtraction of the foreground
emission requires very exacting calibration. In contrast,  we
have followed fairly standard calibration procedures. As such
it seems likely that the discrepancy between our observations and 
existing predictions is probably not genuine; indeed there are
a several purely instrument related  possibilities that 
may account for the discrepancies between our observational 
findings and existing models for the foreground emission. 
We take up first  the issue of calibration error which will 
introduce phase and amplitude errors in the visibilities. 
The fact that the values of $\kappa(U,\Delta \nu)$ are generally
smaller for {\bf data R} as compared to {\bf data I} may be
interpreted as indicating that the visibilities $V(U,\nu)$  are a
combination of two parts, a correlated part which arises 
from for e.g. the effect of calibration errors on discrete sources,
and another whose contribution to  different channels is uncorrelated. 
The ``halos'' that we see around the bright sources is a clear 
indication that calibration problems exist in our data. Phase
errors  which vary with channel would cause decorrelation  
of the visibilities across  different frequencies. Further,
one would expect that the phase errors increase with increasing
baseline length, which is qualitatively consistent with what
we see in Fig.~\ref{fig:r3}. In contrast to the situation
for $\kappa(U,\Delta \nu)$, the contribution from the source subtraction 
residuals  to $V_2(U,0)$ (Fig.~\ref{fig:r2}) can be estimated to 
be small as follows. There are only $\sim 100$ imaging artifacts
with absolute value of flux  $> 20$~mJy ({\bf Data R}, Figure
\ref{fig:a3}) , while about 10,000 such
sources would be needed to  produce the observed visibility 
correlation of $\sim 4\,$Jy$^{2}$ ({\bf Data R}, Figure
\ref{fig:r2}).

The 2D Fourier relation between  the sky brightness and the visibilities
assumed in Section~3 is not strictly valid for GMRT's large
field of view $(\theta_{\rm FWHM}=3.8^{\circ}$). In addition to $u-v$
which are the components of the baseline in the plane normal to the
direction of observation, it is also necessary to consider $w$ the
component along the observing direction. This is a possible source of
error in our visibility correlation analysis. To asses the impact of
the $w$ term we have repeated the analysis using only a limited range
of baselines for which $w\le 100$. We find that limiting the maximum
$w$ value  does not make  any  qualitative change in our
results. The conclusions are unchanged even if we impose $w \le 50$.

 Residual RFI is another  possibility. The visibilities
were clipped at $12 \, {\rm Jy}$ (Section 2.) and this is expected to
remove the strong RFI, but weak RFI contributions will persist in the
data. The RFI electric fields  at  any two antennas  is correlated 
with a time delay $\tau$ which  depends on  position of the RFI source
relative to the antennas and the direction of observation. The RFI 
contribution behaves like the system noise if $\tau$ is greater than 
$\tau_c$ the coherence time of the RFI signal. In this case the RFI 
effectively increases $\sigma$ the rms. fluctuations of the visibilities.
This only changes the error estimates,  and does not affect the 
expected visibility correlations.  RFI sources for which  $\tau <\tau_c$ 
are expected to affect the visibility correlations. This contribution 
will depend on the distribution of the time delays $\tau$s and the
frequency spectrum of the  RFI sources. The analysis of this is beyond 
the scope of this paper. Work is currently underway at the GMRT to
implement more sophisticated real time as well as offline RFI mitigation
schemes.  Future  observations will help assess the improvement that
these schemes as well as better calibration procedures make on
the problem of foreground subtraction. Polarization leakage is another 
important  issue that we plan to take up in future work.  

\section{Acknowledgment}
SSA and SB  would like to thank Prasun Dutta and Kanan K. Datta for their help.  The data
used in this paper were  obtained using GMRT. The
GMRT is run by the National Centre for Radio Astrophysics of the
Tata Institute of Fundamental Research. We thank the GMRT staff for
making these observations possible.

\end{document}